\documentclass[12pt,a4paper,final]{iopart}

\usepackage{iopams}  
\usepackage{graphicx}
\usepackage[breaklinks=true,colorlinks=true,linkcolor=blue,urlcolor=blue,citecolor=blue]{hyperref}

\newcommand{\be}{\begin{equation}}
\newcommand{\ee}{\end{equation}}
\newcommand{\eq}[1]{(\ref{#1})}
\newcommand{\fig}[1]{figure~\ref{#1}}
\def\beq{\begin{eqnarray}}
\def\eeq{\end{eqnarray}}\def\beqa{\begin{eqnarray}}
\def\eeqa{\end{eqnarray}}

\def\bra{\langle}
\def\ket{\rangle}
\def\vq{{\bf q}}

\begin{document}

\title[Particle-hole asymmetry in doped Mott insulators]{Strong particle-hole asymmetry of charge instabilities in doped Mott insulators}

\author{Mat\'{\i}as Bejas$^1$, Andr\'es Greco$^1$ and Hiroyuki Yamase$^{2,3}$}
\address{$^1$Facultad de Ciencias Exactas, Ingenier\'{\i}a y Agrimensura and
Instituto de F\'{\i}sica Rosario (UNR-CONICET),
Avenida Pellegrini 250, 2000 Rosario, Argentina}
\address{$^2$National Institute for Materials Science, Tsukuba 305-0047, Japan}
\address{$^3$Max-Planck-Institute for Solid State Research, D-70569 Stuttgart, Germany}
\eads{\mailto{bejas@ifir-conicet.gov.ar},
      \mailto{agreco@fceia.unr.edu.ar} and
      \mailto{yamase.hiroyuki@nims.go.jp}}

%

\begin{abstract}
We study possible charge instabilities in doped Mott insulators by employing 
the two-dimensional $t$-$J$ model with a positive value of the next 
nearest-neighbor hopping integral $t'$ on a square lattice, which is applicable 
to electron-doped cuprates. 
Although the $d$-wave charge density wave (flux phase) and $d$-wave Pomeranchuk 
instability (nematic order) are dominant instabilities for a negative $t'$ that corresponds 
to hole-doped cuprates, we find that those instabilities 
are strongly suppressed and become relevant only rather close to half filling. 
Instead, various types of bond orders with modulation vectors close to 
$(\pi,\pi)$ are dominant in a moderate doping region.  
Phase separation is also enhanced, but it can be suppressed substantially by 
the nearest-neighbor Coulomb repulsion without affecting the aforementioned charge instabilities.

\end{abstract}


\vspace{2pc}

\submitto{\NJP}

\section{Introduction} \label{sec:intro}

High-temperature cuprate superconductors are 
realized by carrier doping into antiferromagnetic Mott insulators, 
and superconductivity is characterized by $d$-wave symmetry. 
The cuprate superconductors are layered materials; and the
electronic properties in the CuO$_{2}$ plane, where 
Cu sites form a square lattice, hold the key to 
high-temperature superconductivity. 
Its essential physics is believed to be contained in the two-dimensional 
$t$-$J$ and Hubbard models on a square lattice \cite{anderson87,fczhang88}.
Despite these common views, the underlying physics of cuprate 
superconductivity remains highly elusive. 

One of the notorious puzzles in hole-doped cuprates ($h$-cuprates) concerns the 
pseudogap (PG) \cite{timusk99,norman05},  
a gap-like feature in the normal phase even far above the
superconducting onset temperature ($T_{sc}$).
There are two major scenarios for the origin of the PG.
One scenario invokes fluctuations of Cooper pairs above $T_{sc}$ \cite{emery95,norman07,mishra14} 
whereas the other invokes some order competing with superconductivity. 
Recent angle-resolved photoemission spectroscopy \cite{tanaka06,vishik10,kondo11,yoshida12} 
observes the two-gap feature in the electronic band dispersion, in favor of the latter scenario for 
the PG. 
However, it is a matter of considerable debate what kind of order actually develops in the PG. 
The so-called YRZ model \cite{yang06} exploits the concept of the 
resonating-valence-bond theory \cite{anderson04,lee06} and 
successfully captures some features of the PG. 
On the other hand, 
various experimental observations in the PG state are also well captured 
in terms of charge instabilities such as 
$d$-wave charge density wave ($d$CDW) \cite{chakravarty01,cappelluti99,greco09,bejas11,greco11},  
a loop current order \cite{varma99,varma06a}, 
$d$-wave Pomeranchuk instability ($d$PI) \cite{yamase09,hackl09,yamase12},   
conventional charge density wave (CDW) \cite{castellani95,becca96,hashimoto10,he11}   
including stripes \cite{kivelson03,vojta09},  and 
phase separation (PS) \cite{castellani95,becca96,emery93}.

Quite recently a charge-order instability was observed by X-rays
in two different $h$-cuprates, Y-based~\cite{wu11,ghiringhelli12,chang12} and 
Bi-based~\cite{comin14,neto14} cuprates. 
This charge order is not accompanied by a magnetic order, 
in sharp contrast with the spin-charge stripes \cite{tranquada95} discussed 
extensively in La-based cuprates \cite{kivelson03}.
Thus, charge-order instabilities in cuprates have attracted renewed interest.
A comprehensive study \cite{bejas12} about possible charge orders in the $t$-$J$ model 
showed that doped Mott insulators exhibit strong tendencies toward 
the $d$CDW and $d$PI.
In particular, the incommensurate $d$PI \cite{bejas12,metlitski10,metlitski10a,holder12,husemann12,efetov13,bulut13,sachdev13} 
attracts much interest. 
However, its modulation vector is not consistent with the experiments \cite{wu11,ghiringhelli12,chang12,comin14,neto14},
requiring a further study both theoretically and experimentally. 

Electrons can also be doped into the parent compound of cuprates. 
Electron-doped cuprates ($e$-cuprates) \cite{armitage10},
however, look very different from $h$-cuprates. 
A PG similar to that found in $h$-cuprates is not clearly observed. 
If the PG indeed originates from some charge order as discussed 
regarding $h$-cuprates, it seems natural to assume that charge-order tendencies 
are strongly suppressed in $e$-cuprates. 
On the other hand, a recent finding of collective excitations in optimal 
$e$-cuprates \cite{wslee13} suggests that 
a charge-order tendency can be present.

Electron-cuprates cuprates have often been discussed via a comparison with $h$-cuprates, 
focusing on specific aspects, e.g., pairing properties \cite{white99,martins01},
magnetic properties \cite{tohyama94,gooding94},
stability of charge stripes but with different conclusions \cite{white99,tohyama99},
and optical conductivity \cite{tohyama01}.
A recent comprehensive study using variational Monte Carlo \cite{yokoyama13}  
showed that superconductivity is enhanced but antiferromagnetism 
is suppressed in $h$-cuprates, whereas the opposite occurs in $e$-cuprates, 
nicely demonstrating the experimental fact. 
In spite of these works, charge-order tendencies in $e$-cuprates 
have not been clarified. 

In this paper,
we study all possible charge instabilities 
in $e$-cuprates in the framework of the two-dimensional $t$-$t'$-$J$ model.
We employ a similar theoretical framework in which charge-order tendencies have been studied 
comprehensively for $h$-cuprates \cite{bejas12}. In this sense, the current work 
is a complement to \cite{bejas12} and is expected to clarify charge-order tendencies 
in $e$-cuprates in the most comprehensive way through a comparison with those in $h$-cuprates. 
We find that charge-order tendencies 
exhibit a very strong particle-hole asymmetry. 
Although $h$-cuprates have strong tendencies toward the $d$CDW and $d$PI \cite{bejas12},
these orders are substantially suppressed in $e$-cuprates. 
Instead, various bond orders with large momenta near $(\pi,\pi)$ 
are favored in a moderated doping region. 
In section~\ref{sec:model}, we define our model and explain our methods. 
Numerical results are presented in section~\ref{sec:results}. 
We discuss possible charge instabilities and the PG in $e$-cuprates in section~\ref{sec:discussions}, 
followed by conclusions in section~\ref{sec:conclusions}. 

\section{Model and formalism} \label{sec:model}
We study charge instabilities in the two-dimensional $t$-$t'$-$J$ model by 
including the nearest-neighbor Coulomb interaction $V$, 
\be\label{Hc}
H = -\sum_{i, j,\sigma} t_{i j}\tilde{c}^\dag_{i\sigma}
\tilde{c}_{j\sigma}
+ J \sum_{\bra i,\, j\ket} \left( \vec{S}_i \cdot \vec{S}_j-\frac{1}{4} n_i n_j \right)
+ V \sum_{\bra i,\, j\ket} n_i n_j \, .
\ee
$t_{i j} = t$ $(t')$ is the hopping integral between the first (second) nearest-neighbor 
sites on a square lattice;  $J$ and $V$ are the exchange interaction
and the Coulomb repulsion, respectively, between the nearest-neighbor sites. 
$\bra i, j\ket$ indicates a nearest-neighbor pair. 
$\tilde{c}^\dag_{i\sigma}$ and $\tilde{c}_{i\sigma}$ are 
the creation and annihilation operators of electrons 
with spin $\sigma$ ($\sigma = \downarrow$,$\uparrow$),  respectively, 
in Fock space without any double occupancy. 
$n_i=\sum_{\sigma} \tilde{c}^\dag_{i\sigma}\tilde{c}_{i\sigma}$ 
is the electron density operator and $\vec{S}_i$ is the spin operator 
in that space. Although the $V$ term is usually neglected in the analysis of the $t$-$J$ model, 
its presence is natural, as seen in the derivation of the $t$-$J$ model 
from the generalized Hubbard model \cite{chao77}.
We found that a role of the $V$ term in the current study is to suppress 
the strong tendency toward PS (see section \ref{sec:ps}) and does not affect the charge instabilities 
originating from the $J$ term (see section \ref{sec:ciJ}). 
There are higher-order corrections to the t-J model, such as correlated
hopping terms \cite{chao77,feiner96}.
We expect that those corrections do not blur our principal physics
originating from the $J$ term as well as the strong correlation effect 
contained in (\ref{Hc}). We thus discard them in the current study.

We study the Hamiltonian (\ref{Hc}) in a large-$N$ technique formulated in a path integral 
representation of the Hubbard $X$ operators \cite{bejas12,foussats04}.
Because details of the formalism were presented in \cite{bejas12}, 
we provide a simple sketch of that here. 

We first write the Hamiltonian (\ref{Hc}) in terms of Hubbard operators \cite{hubbard63} 
via $\tilde{c}^\dag_{i \sigma}=X_i^{\sigma 0}$, 
$\tilde{c}_{i \sigma}=X_i^{0 \sigma}$, 
$S_i^+=X_i^{\uparrow \downarrow}$, 
$S_i^-=X_i^{\downarrow \uparrow}$, 
$S_i^{z}=(X_i^{\uparrow \uparrow}-X_i^{\downarrow \downarrow})/2$, 
and $n_i=X_i^{\uparrow \uparrow}+X_i^{\downarrow \downarrow}$;  
 $X_i^{00}$ will also be introduced later [see \eq{eq:v1}]. 
 We then extend the spin degree of freedom to $N$ channels and obtain 
the Hamiltonian in the large-$N$ formalism, 
\begin{eqnarray}
 H_{N} &=& - \frac{1}{N}\sum_{i, j, p}\; t_{i j} X_{i}^{p 0}X_{j}^{0p} 
+ \frac{J}{2N} \sum_{\bra i,j\ket, pp'} \left( X_{i}^{p p'}X_{j}^{p' p} - X_{i}^{p p} X_{j}^{p' p'} \right) \nonumber \\
&&+ \frac{V}{N}\sum_{\bra i,j\ket, p p'} X_{i}^{p p} X_{j}^{p' p'}
-\mu\sum_{i,p}\;X_{i}^{p p} \label{eq:H} \,.
\end{eqnarray}
The spin index $\sigma$ is extended to a new index $p$, which 
runs from $1$ to $N$. To obtain a finite theory in the $N$-infinite limit,  
$t$, $t'$, $J$ and $V$ are rescaled as
$t/N$, $t'/N$, $J/N$ and $V/N$, respectively. 
The chemical potential $\mu$ is introduced in \eq{eq:H}. 

The Hamiltonian (\ref{eq:H}) can be formulated in a path integral representation \cite{foussats04}.
Our Euclidean Lagrangian then reads 
\begin{eqnarray}
L_E =  \frac{1}{2}
\sum_{i, p}\frac{\left( {\dot{X_{i}}}^{0 p}\;X_{i}^{p 0} + {\dot{X_{i}}}^{p 0}\;
X_{i}^{0 p} \right)} {X_{i}^{0 0}} + H_{N}
\label{lagrangian}
\end{eqnarray}
with the following two additional constraints, 
\begin{eqnarray}
X_{i}^{0 0} + \sum_{p} X_{i}^{p p} - \frac{N}{2}=0 \; , \label{eq:v1}
\end{eqnarray}
and
\begin{eqnarray}
X_{i}^{p p'} - \frac{X_{i}^{p 0} X_{i}^{0p'}}{X_{i}^{0 0}}=
0 \;, \label{eq:v2}
\end{eqnarray}
which are imposed on the path integral via two $\delta$-functions.
In \eq{lagrangian}, ${\dot{X_{i}}}^{p 0}=\partial_\tau {X_{i}}^{p 0}$ 
and $\tau$ is the Euclidean time, namely $\tau={\rm i}t$.

We first write $X_{i}^{p p}$ in the Hamiltonian (\ref{eq:H}) 
in terms of $X_{i}^{0 0}$ by using \eq{eq:v1}. This 
ensures that the $V$ term vanishes at half filling due to 
strong correlation effects in the Mott insulator.   
The completeness condition \eq{eq:v1} imposed by the $\delta$-function is now
described by introducing Lagrange multipliers $\lambda_{i}$.
We describe $X_{i}^{0 0}$ and $\lambda_i$  in terms of
static mean-field values, $r_0$ and $\lambda_0$, and fluctuation fields, 
$\delta R_i$ and $\delta \lambda_i$:
\begin{eqnarray}
X_{i}^{0 0} &=& N r_{0}(1 + \delta R_{i}) \nonumber \\
\lambda_{i} &=&\lambda_{0}+ \delta{\lambda_{i}} \,.
\label{lambdai}
\end{eqnarray}
From the completeness condition \eq{eq:v1}, $r_0$ is equal to $\delta/2$,  
where $\delta$ is the doping rate away from half filling. 
We then eliminate $X^{p p'}$ by implementing the $\delta$-function associated with \eq{eq:v2}. 
This procedure creates interaction terms such as $X_{i}^{p 0}X_{i}^{0 p'} X_{j}^{p' 0} X_{j}^{0 p}$, 
which are decoupled through a Hubbard-Stratonovich transformation 
by introducing a field associated with a bond variable, 
\be
\Delta_{ij} = \frac{J}{N r_0} \sum_{p} \frac{X_{j}^{p 0} X_{i}^{0 p}}{ \sqrt{(1 + \delta R_{i})
(1 + \delta R_{j})}} \,. 
\ee
The field $\Delta_{ij}$ is parameterized by 
\be
\Delta_i^{\eta}=\Delta (1+ r_i^\eta+\rmi A_i^\eta)\,,
\label{staticDelta}
\ee  
where $r_i^{\eta}$ and $A_i^{\eta}$ correspond to the real and imaginary parts of the 
fluctuations of the bond variable,
respectively, and $\Delta$ is a static mean-field value. 
The index $\eta$ takes two values associated with the bond directions
 ${\eta}_{1}=(1,0)$ and ${\eta}_{2}=(0,1)$ on a square lattice. 
After expanding $1/(1+\delta R)$ in powers of $\delta R$, 
we obtain an effective Lagrangian, which can be written in terms of 
a six-component bosonic field 
\be
\delta X^{a} = (\delta
R\;,\;\delta{\lambda},\; r^{{\eta}_{1}},\;r^{{\eta}_{2}}
,\; A^{{\eta}_{1}},\;
A^{{\eta}_{2}})\,,
\label{6-boson}
\ee 
the fermionic fields $X^{0 p}$ and $X^{p 0}$, and their interactions. 
Any physical quantity can then be calculated at a given order 
by counting powers of $1/N$ in a corresponding Feynman diagram, 
providing a controllable scheme. 
The Feynman rules are given in figure~1 in \cite{foussats04}. 

Because the bosonic field has six components [see \eq{6-boson}], 
its bare propagator $D^{(0)}_{ab}(\vq,\mathrm{i}\omega_n)$ 
is given by a $6\times6$ matrix; 
${\vq}$ and ${\rm i}\omega_{n}$ are the momentum and bosonic 
Matsubara frequency, respectively. 
The quantity $D^{(0)}_{ab}(\vq,\mathrm{i}\omega_n)$ 
describes all possible types of bare charge susceptibilities. 
From the Dyson equation the dressed propagator is given by 
\be
D^{-1}_{ab}(\vq,\mathrm{i}\omega_n)
= \left[ D^{(0)}_{ab}(\vq,\mathrm{i}\omega_n) \right]^{-1} - \Pi_{ab}(\vq,\mathrm{i}\omega_n)\,.
\label{dyson}
\ee
The bosonic propagator acquires the self-energy 
$\Pi_{ab}(\vq,\mathrm{i}\omega_n)$ already at the leading order 
[see equations (15)-(18) in \cite{bejas12} 
for the explicit expression of $D^{-1}_{ab}(\vq,\mathrm{i}\omega_n)$].  
As a result, an eigenvalue of $D_{ab}(\vq,0)$ can diverge, 
leading to a charge instability with a modulation vector $\vq$. 

From the $N$-extended 
completeness condition \eq{eq:v1}, we see that the charge operator 
$X^{00}$ is $O(N)$, whereas the operators $X^{pp}$ are $O(1)$. 
Consequently, the $1/N$ approach emphasizes   
the effective charge interactions. 
In fact, collective effects enter the spin susceptibilities in the next-to-leading order. 
This is also the case for superconductivity \cite{cappelluti99}.
Hence instabilities of the paramagnetic phase 
are expected only, in the leading order, in the charge sector. 
This is an advantage of our method and allows us 
to explore all possible charge instabilities exclusively. 
We therefore retain our approximation at the leading order. 
In the leading order theory, however, we cannot address the ground state, 
which likely exhibits superconductivity. 
Hence our results  should be interpreted as microscopic indications 
of what kind of charge instabilities  
become relevant in a parameter region where magnetism and 
superconductivity are absent. 

In the leading order, our formalism agrees with the $1/N$ slave-boson 
formalism \cite{morse91} as well as results in another formalism of 
the $1/N$ expansion \cite{cappelluti99}.
Our formalism was also verified, in the next-to-leading order, to yield results consistent with 
the exact diagonalization \cite{merino03,bejas06}.
In the next section we will also pay attention to the consistency between 
our results and existing literature. 

\section{Results}\label{sec:results}
The $t$-$t'$-$J$ model with $t'<0$ has been extensively studied 
in the context of $h$-cuprates. Because the model is defined 
in Fock space without any double occupancy, 
we perform a particle-hole transformation \cite{tohyama94} 
for studying $e$-cuprates. 
This is implemented by taking a positive value of $t'$ \cite{white99,martins01,tohyama94,gooding94,tohyama99,tohyama01}.

In the following we set $t=1$ and all quantities with the dimension of energy are in 
units of $t$ except for \fig{fig6}. A typical value of $t$ in cuprates is estimated to be around 
500 meV \cite{hybertsen90}.

\subsection{Possible charge instabilities in $e$-cuprates}\label{sec:ci}
\begin{figure}
\begin{center}
\setlength{\unitlength}{1cm}
\includegraphics[width=8cm,angle=0]{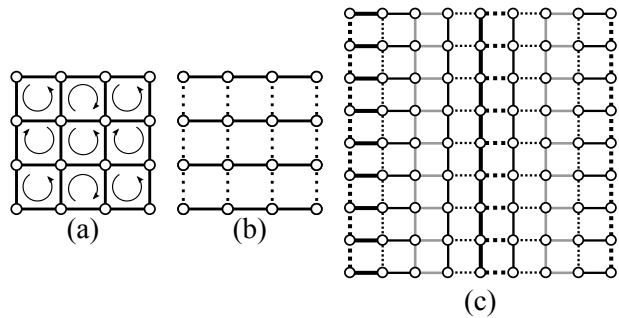}
\end{center}
\caption{Sketch of (a) $d$CDW with $\vq=(\pi,\pi)$, (b) $d$PI with $\vq=(0,0)$,   
and (c) $d$PI with $\vq=(\pi/4,0)$. 
The black lines in (b) and (c) denote a stronger (solid line) and weaker (dotted line) 
bond relative to the mean-field bond variable (gray line), namely 
the constant term on the right-hand side of \eq{staticDelta}. 
The width of the lines in (c) indicates the modulation amplitude.
}
\label{fig1}
\end{figure}

We compute the static charge susceptibilities $D_{ab}(\vq,0)$ from \eq{dyson}, 
which are given by a $6 \times 6$ matrix, 
at the leading order of the large-$N$ expansion. 
When an eigenvalue of the inverse of the matrix, namely $D^{-1}_{ab}(\vq,0)$, 
crosses zero at a given doping rate $\delta$, temperature $T$, and $\vq$, 
a charge instability with a modulation vector $\vq$ occurs and the ordering pattern 
is determined by the corresponding eigenvector $V^a$. 
The eigenvectors that we have found are the same as those 
in \cite{bejas12}, although we employ the opposite sign of $t'$ here. 
We explain these eigenvectors one by one in the following paragraphs. 

i) $V^a \propto (0,0,0,0,1,-1)$, which corresponds to 
the $d$CDW (flux phase) 
with $\vq=(\pi,\pi)$ \cite{cappelluti99,foussats04,morse91,affleck88a,affleck89}. 
In this phase, currents flow in each plaquette as shown in \fig{fig1} (a). 

ii) $V^a \propto (0,0,1,-1,0,0)$, which corresponds to a $d$PI 
with $\vq =(0,0)$ (commensurate) \cite{yamase00a,yamase00b,metzner00}
or close to it (incommensurate) \cite{bejas12,metlitski10,metlitski10a,holder12,husemann12,efetov13,bulut13,sachdev13}.
The $d$PI leads to the electronic nematic state as an instability of the 
paramagnetic state. The commensurate and the incommensurate $d$PI 
are shown in figures~\ref{fig1}(b) and (c), respectively. 

\begin{figure}
\begin{center}
\setlength{\unitlength}{1cm}
\includegraphics[width=8cm,angle=0]{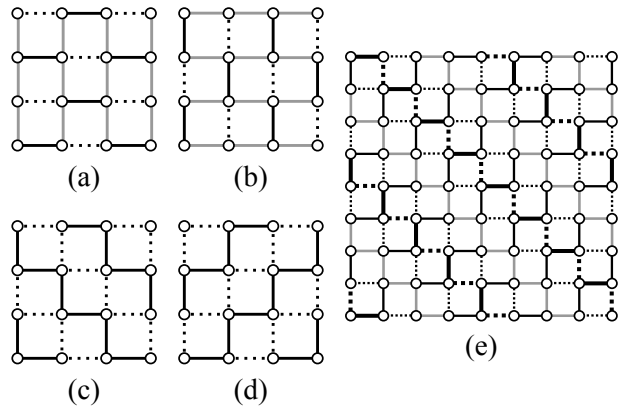}
\end{center}
\caption{Sketch of (a) BOP$_{x}$, (b) BOP$_{y}$, (c) BOP$_{xy}$, and (d) BOP$_{x\bar{y}}$  
for $\vq=(\pi,\pi)$; (e) BOP$_{x\bar{y}}$ with $\vq=(3\pi/4,3\pi/4)$. 
Gray, solid, and dotted lines are explained in \fig{fig1}.
}
\label{fig2}
\end{figure}

iii) $V^a \propto (0,0,1,0,0,0), (0,0,0,1,0,0), (0,0,1,1,0,0)$, and $(0,0,1,-1,0,0)$, 
which correspond to the bond-order phase (BOP) \cite{cappelluti99,foussats04,morse91} 
with $\vq=(\pi,\pi)$ or close to it, 
with four different patterns: BOP$_{x}$, BOP$_{y}$, BOP$_{xy}$, and BOP$_{x\bar{y}}$, 
respectively [see figures~\ref{fig2}(a)-(d)]. 
BOP$_{x(y)}$ is a phase that has a bond amplitude modulated only 
along the $x(y)$ direction, 
whereas BOP$_{xy(x\bar{y})}$ with $(\pi,\pi)$ 
has a bond amplitude modulated along both the $x$ and $y$ directions, and its relative phase 
is inphase (antiphase). Because the $d$PI and BOP$_{x\bar{y}}$ belong to the same 
eigenvector $(0,0,1,-1,0,0)$, the $d$PI with $\vq \approx (\pi,\pi)$ is 
equivalent to the BOP$_{x\bar{y}}$. However, 
the term of the $d$PI makes sense only for a small $\vq$, 
and thus we use the term BOP$_{x\bar{y}}$ when $\vq$ is no longer close to $(0,0)$. 
We also sketch BOP$_{x\bar{y}}$ with $\vq=(3\pi/4,3\pi/4)$ in \fig{fig2}(e).   
Such an order can occur for a large $t'$ (see \fig{fig5}). 

iv) $V^a \propto (1,0,0,0,0,0)$, which corresponds to a PS with $\vq=(0,0)$. 
Conventional CDW including charge stripes 
also belongs to the same eigenvector, but with a finite $\vq$. 
Such an instability was not found in the current study. 
This work is in favor of \cite{white99} more 
than \cite{tohyama99} regarding the stability of charge stripes.

In the following, we will specify a parameter region where each charge instability 
can occur by varying doping rate $\delta$, temperature $T$, 
and the next-nearest neighbor hopping $t'$. 
Because we determine critical lines of charge instabilities by studying 
susceptibility, the transition is always continuous and a possible first-order 
transition is not considered in the current study. 

Before presenting our results, we emphasize that 
our general susceptibility \eq{dyson} 
considers all possible charge instabilities. As mentioned in the Introduction, 
various charge instabilities 
are discussed in the context of the PG, and most of them are indeed found in this work 
except for conventional CDW including stripes; 
the loop current order is beyond the scope of the one-band $t$-$J$ model.  
A bond-order modulated flux phase was discussed in variational Monte Carlo in the $t$-$J$ model 
at zero temperature \cite{poilblanc05,weber06}. 
Such a state is described by the mixture of two eigenvectors, $(0,0,0,0,1,-1)$ and $(0,0,1,1,0,0)$,  
in the current theory, but is not found here. This suggests that 
such a state may not occur 
as an instability from the normal phase, but may occur 
as an additional instability inside the symmetry broken phase 
characterized by either eigenvector. 
This possibility cannot be addressed in the current theory because 
we perform the stability analysis 
of the normal phase in terms of the susceptibility.

\subsection{Phase separation}\label{sec:ps}
\begin{figure}
\begin{center}
\setlength{\unitlength}{1cm}
\includegraphics[width=8cm,angle=0]{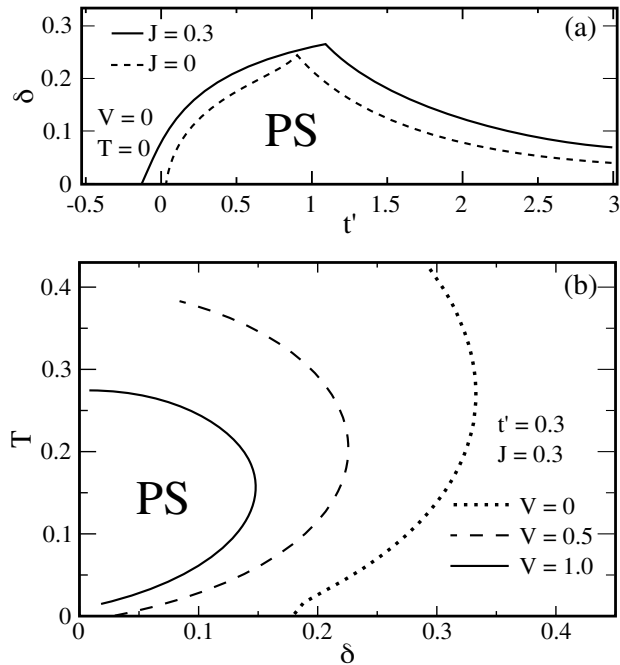}
\end{center}
\caption{
(a) Phase separation in the plane of $t'$ and $\delta$ for $J=0$ and $0.3$ 
at $V=T=0$.
(b) Phase separation in the plane of $\delta$ and $T$ for 
various $V$ at $J=0.3$ and $t'=0.3$.
}
\label{fig3}
\end{figure}

We first discuss PS.  
As seen in the literature \cite{martins01,gooding94,macridin06},
PS is strongly enhanced for a positive $t'$. 
Figure~\ref{fig3}(a) shows PS in the plane of $t'$ and $\delta$ 
at $V=T=0$ for $J=0$ and $0.3$; $\delta=0$ corresponds to half filling, and 
$\delta$ denotes the electron (hole) doping rate for $t'>0 (<0)$. 
As seen from the large slope at $t'\approx -0.1$ for $J=0.3$,  
PS is rapidly stabilized with increasing $t' (> -0.1)$ and 
extends to 15\% doping already around $t' \approx 0.1$. 
PS is monotonically enhanced up to $t' \approx 1$ and is suppressed for $t'>1$.
A similar result was also obtained by exact diagonalization \cite{martins01}.
Although PS is enhanced by the $J$ term, 
the $t'$ dependence of PS is well captured by the result of $J=0$. 
Although one might assume that a finite $J$ is necessary to obtain PS, 
the kinetic term in the $t$-$J$ model [first term in the Hamiltonian (\ref{Hc}) and (\ref{eq:H})]   
is not a usual non-interacting term but already contains strong 
correlation effects coming from the local constraints (\ref{eq:v1}) and (\ref{eq:v2}). 
Figure~\ref{fig3} thus clearly demonstrates that PS originates from strong correlation effects, 
in line with the result obtained by the dynamical cluster approximation in the 
strong coupling Hubbard model \cite{macridin06}.

Figure~\ref{fig3}(b) shows the region of PS in the plane of $\delta$ and $T$ for several 
choices of $V$ at $J=0.3$ and $t'=0.3$. 
PS occurs on the left side of the critical line.   
As expected, PS is substantially suppressed 
by increasing $V$. We verified that no additional CDW instability was 
triggered for the current values of $V$. 
The doping region of PS shrinks at low and high $T$. 
Because of such reentrant behavior, PS can be stabilized at a finite $T$ 
even if it does not occur at $T=0$. 
A result similar to \fig{fig3}(b) 
was also obtained in the Hubbard model in strong coupling \cite{koch04}.
We, however, note that our PS is not a pure PS especially for high $T$.
Although the eigenvector of PS contains the component of $(1,0,0,0,0,0)$ 
almost 100\% close to zero temperature, the weight from other components, 
especially from $(0,0,1,1,0,0)$, increases with increasing $T$. 
For example, the weight of the $(1,0,0,0,0,0)$ component  
is reduced to about 80\% (50\%) at $T=0.1 (0.2)$ for $V=1.0$; 
such a reduction occurs at a higher temperature 
for a smaller $V$.  

Given that PS tendencies found here are consistent with results obtained in 
other methods \cite{martins01,gooding94,macridin06,koch04} 
we believe that PS is a genuine feature of the $t$-$J$ model. 
However, when PS occurs, charge accumulates in one region more than in the other region. 
In this case, it is readily expected that long-range Coulomb 
interaction, which is not considered in the $t$-$J$ model, 
may stabilize an inhomogeneous state. This possibility is worth exploring further.  

\subsection{Charge instabilities from the $J$ term}\label{sec:ciJ}
We now study all the other possible charge instabilities and fix 
$J=0.3$ which is believed to be appropriate to 
cuprates \cite{tohyama94,gooding94,hybertsen90}.
We choose $V=1$ to suppress PS, but  
charge instabilities from the $J$ term turn out to be rather 
insensitive to the choice of $V$. 
The latter aspect of the $V$ term might be surprising from a view of weak coupling theory. 
However, as seen in our formalism in section~\ref{sec:model}, 
the $V$ term vanishes at half filling and 
the current theory belongs to a strong coupling theory formulated 
in terms of the Hubbard $X$ operators. 
The value of $t'$ is estimated to be around $t'=0.2 \sim 0.4$ 
for $e$-cuprates \cite{tohyama94,gooding94,hybertsen90}.
Because a realistic value of $t$ is around 500 meV in cuprates \cite{hybertsen90},     
the temperature range we are interested in is below $T=0.04$-$0.02$, 
which corresponds to a region below 200-100 K. 

\begin{figure}
\begin{center}
\setlength{\unitlength}{1cm}
\includegraphics[width=7cm,angle=0]{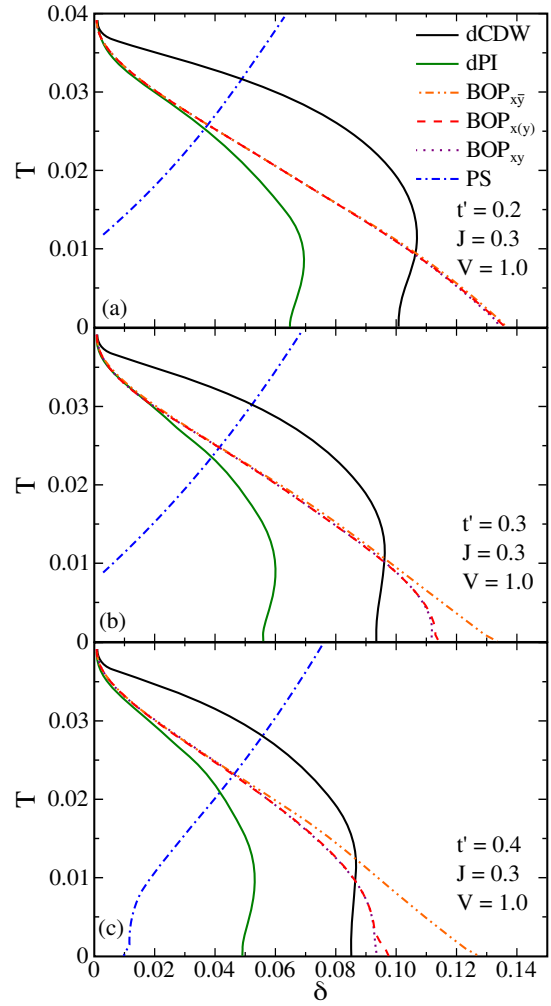}
\end{center}
\caption{
Doping dependence of critical temperatures $T_c$ of 
$d$CDW, $d$PI, BOP$_{x\bar{y}}$, BOP$_{x(y)}$, BOP$_{xy}$, and PS for $J=0.3$ and 
$V=1$; (a) $t'=0.2$, (b) $0.3$, and (c) $0.4$. 
The instability occurs below the corresponding critical line  
except for PS, which is stabilized on the left side of the critical line. 
}
\label{fig4}
\end{figure}

Figure \ref{fig4} shows a phase diagram in the plane of $\delta$ and $T$. 
It extends the information we obtained in figure 2 of 
\cite{bejas12} for $h$-cuprates by showing the effect of a positive $t'$ appropriate 
for $e$-cuprates. 
As already seen in \fig{fig3}(b), PS occurs on the side of half filling and is enhanced 
at high $T$.
In contrast with PS, other charge instabilities are driven by 
the $J$ term. In \fig{fig4}(a), they occur at lower temperatures ($0< T < 0.04$) below $\delta \approx 0.14$.
This region is actually what we are interested in, in the context of $e$-cuprates. 
We obtain three different types of charge instabilities: 
$d$CDW with $\vq=(\pi,\pi)$, $d$PI with $\vq=(0,0)$, and various BOPs such as 
BOP$_{x(y)}$, BOP$_{xy}$, and BOP$_{x\bar{y}}$ with $\vq\approx (\pi,\pi)$. 
The $d$PI is suppressed most strongly and is stabilized only 
rather close to half filling.  
Although the $d$CDW is the leading instability in $\delta \lesssim 0.1$,  
BOP$_{x(y)}$, BOP$_{xy}$, and BOP$_{x\bar{y}}$ become dominant 
in the region $0.1 \leq \delta \leq 0.14$ and 
show instabilities almost simultaneously. 
As $t'$ increases, charge instabilities except for PS are suppressed 
and stabilized closer to half filling, as seen in figures~\ref{fig4}(b) and (c). 
Among various BOPs, BOP$_{x\bar{y}}$ becomes the leading instability at low $T$
in a moderate doping region with increasing $t'$. 
We verified that the results of \fig{fig4}, except for PS, 
do not depend on a precise choice of the value of $V$. 
Because we compute the general susceptibility in the paramagnetic state, 
\fig{fig4} should be interpreted as a hierarchy of different charge instabilities. 
For instance, in \fig{fig4}(c) at $\delta=0.08$, 
$d$CDW and BOP$_{x\bar{y}}$ are the leading and the next-to-leading instabilities, 
respectively. BOP$_{x(y)}$ and BOP$_{xy}$ are degenerate and the third-to-leading instability.

\begin{figure}[t]
\begin{center}
\setlength{\unitlength}{1cm}
\includegraphics[width=10cm,angle=0]{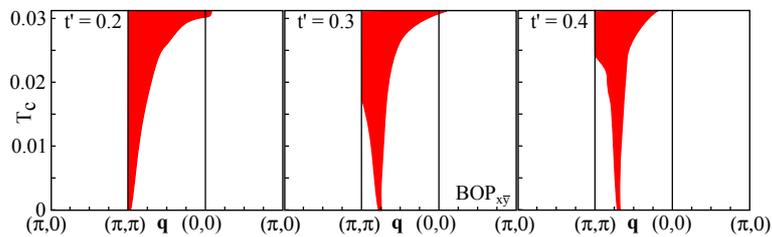}
\end{center}
\caption{
Modulation vectors of BOP$_{x\bar{y}}$ 
along the corresponding critical lines in \fig{fig4}. 
}
\label{fig5}
\end{figure}

\begin{figure}[t]
\begin{center}
\setlength{\unitlength}{1cm}
\includegraphics[width=13cm,angle=0]{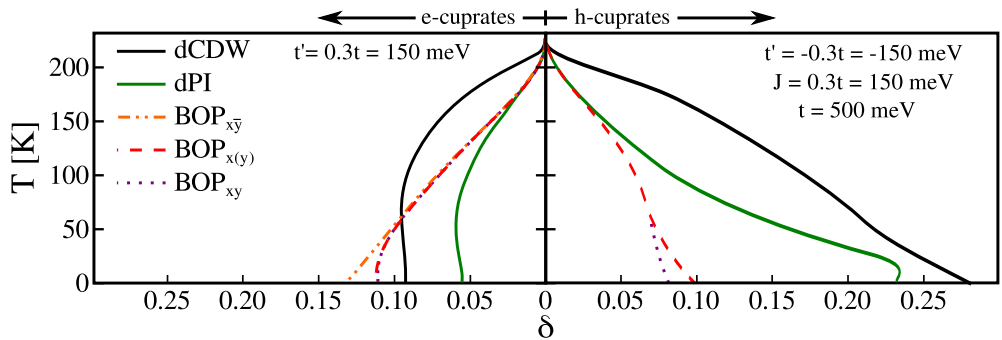}
\end{center}
\caption{
Comparison between the current results for $e$-cuprates (left panel), which is the same as \fig{fig4}(b), 
and the previous results \cite{bejas12} for $h$-cuprates (right panel). 
Using $t = 500$ meV, the temperature scale is given in K for an easy comparison with experiments. 
}
\label{fig6}
\end{figure}

Although the $d$PI  and $d$CDW instabilities always occur at 
$\vq=(0,0)$ and $(\pi,\pi)$, respectively, the modulation vectors  
of BOP$_{xy}$ and BOP$_{x(y)}$ show the instabilities at $\vq=(\pi,\pi)$ and 
shift at very low $T (\lesssim 0.005)$ slightly toward the direction $(\pi,\pi)$-$(\pi,0)$ 
for BOP$_{xy}$ and BOP$_{x}$ and the direction $(\pi,\pi)$-$(0,\pi)$ for BOP$_{y}$. 
The modulation vector of BOP$_{x\bar{y}}$ is shown in \fig{fig5} along 
its critical line in figures~\ref{fig4}(a)-(c); hence the doping rate 
also changes with changing $T_c$. 
In contrast to the case of BOP$_{xy}$ and BOP$_{x(y)}$, 
the charge susceptibility corresponding to BOP$_{x\bar{y}}$ 
is rather flat in momentum space. 
We thus plot the modulation vectors where the inverse of the 
charge susceptibility is less than $10^{-4}$. 
The width of such a $\vq$ region at a fixed temperature indicates how sharp 
the susceptibility is in momentum space. 
The modulation vector does not extend to the side of $(\pi,\pi)$-$(\pi,0)$ direction 
because the eigenvector there changes to BOP$_{x}$. 
As $T$ decreases, the susceptibility becomes sharper. 
The modulation vector then becomes $\vq=(\pi,\pi)$ for $t'=0.2$ 
and shifts toward the diagonal direction along $(\pi,\pi)$-$(0,0)$ for a larger $t'$.

\begin{figure}[t]
\begin{center}
\setlength{\unitlength}{1cm}
\includegraphics[width=12cm,angle=0]{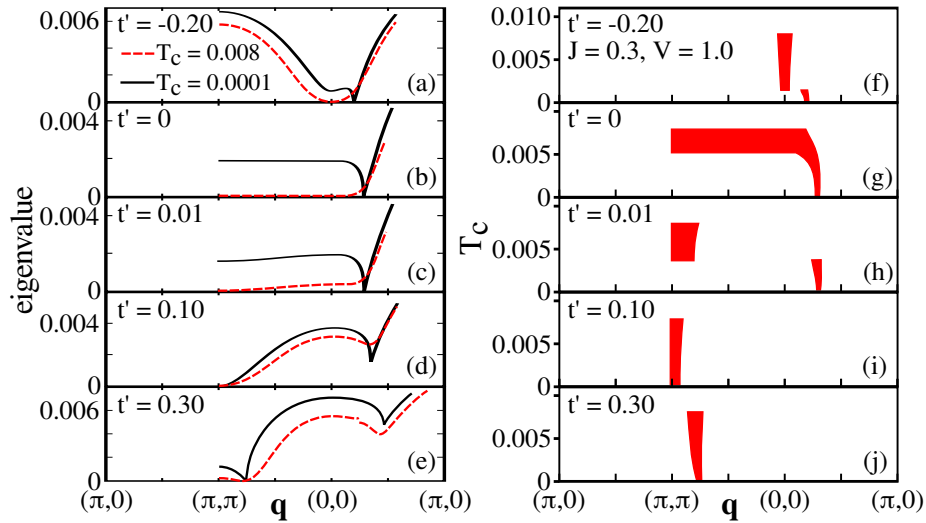}
\end{center}
\caption{
(a)-(e) $\bf{q}$ dependence of the eigenvalue corresponding to 
the eigenvector $(0,0,1,-1,0,0)$ at the critical temperatures 
$T_c=0.008$ and $0.0001$ for a sequence of $t'$. 
The instability occurs when the eigenvalue crosses zero. 
(f)-(j) $T_c$ dependence of the modulation vector $\bf{q}$ 
of the charge instability with the eigenvector $(0,0,1,-1,0,0)$. 
The result (j) is the same as \fig{fig5} for $t'=0.3$ in 
a low temperature region.  
}
\label{fig7}
\end{figure}

Compared with the results for $t'<0$ obtained in \cite{bejas12}, 
charge instabilities, except for PS, show a much weaker 
dependence of $t'$ for $t'>0$. 
To show explicitly the strong particle-hole asymmetry of charge instabilities, 
we compare in \fig{fig6} 
the current results for $t'=0.3$ with our previous results for 
$h$-cuprates obtained in \cite{bejas12} for the same parameter set except for 
the sign of $t'$. 
Although the tendencies of BOPs are even weaker 
than those of the $d$PI and $d$CDW for $t'<0$, 
various BOPs extend to a moderate doping for $t'>0$ and become dominant there. 
Both the $d$CDW and the $d$PI are strongly suppressed for $e$-cuprates compared with 
$h$-cuprates. Whereas the $d$CDW can be still a relevant instability in $e$-cuprates, 
the tendency toward the $d$PI becomes the weakest 
when the sign of $t'$ is reversed. To understand such a drastic change for the $d$PI,
we closely study how the modulation vector of the $d$PI evolves by changing $t'$. 
In the left-hand panels in \fig{fig7}, we show the eigenvalue of the 
inverse of the susceptibility for the eigenvector $(0,0,1,-1,0,0)$ at 
$T_c=0.008$ and $0.0001$ for a sequence of $t'$. 
In the right-hand panels the corresponding modulation vector $\vq$ 
is summarized by determining the momentum $\vq$ at which 
the eigenvalue becomes less than $10^{-4}$ at each temperature. 
For $t'=-0.2$ [figures~\ref{fig7}(a) and (f)] the $d$PI occurs at $\vq=(0,0)$ and 
slightly away from it at very low $T$. 
For $t'=0$  [figures~\ref{fig7}(b) and (g)] 
the susceptibility of the $d$PI becomes flat along the direction 
$(0,0)$-$(\pi,\pi)$, but 
eventually an incommensurate $\vq$ 
is favored along the direction $(0,0)$-$(\pi,0)$ at low $T$. 
The flat feature is a special aspect of the $d$PI susceptibility, 
which becomes exactly flat along the $(0,0)$-$(\pi,\pi)$ direction for any $T$ and $\delta$ 
for $t'=0$ \cite{bejas12}.
With the inclusion of a tiny $t' (=0.01)$ [\fig{fig7}(c)], the flat structure 
is slightly slanted and the eigenvalue at $\vq=(\pi,\pi)$ becomes 
smaller than that at $(0,0)$. As a result, the instability 
occurs at $\vq=(\pi,\pi)$ at high $T$, which is equivalent to BOP$_{x\bar{y}}$. 
Although the flat feature still remains at low $T$ [\fig{fig7}(c)], 
an incommensurate $d$PI develops along the direction of $(0,0)$-$(\pi,0)$, 
similar to the results for $t'=0$. 
A value of $t' \geq 0.10$ is sufficient to completely destroy the $d$PI with a small $\vq$ 
and stabilizes BOP$_{x\bar{y}}$ with $\vq \approx (\pi,\pi)$ in the entire
temperature region [figures~\ref{fig7}(i) and (j)]. 
The eigenvalue \cite{misc-eigenvector}, however, still has a local minimum along 
$(0,0)$-$(\pi,0)$ [figures~\ref{fig7}(d) and (e)]. 
These results, therefore, imply that 
the reason why the stabilization of the $d$PI 
changes rapidly by changing the sign of $t'$ lies in 
the special feature of the $d$PI susceptibility, which exhibits an exactly 
flat structure along $(0,0)$-$(\pi,\pi)$ direction for $t'=0$.

\section {Discussions} \label{sec:discussions}
The $e$-cuprates are characterized by a positive $t'$ in the $t$-$t'$-$J$ model \cite{white99,martins01,tohyama94,gooding94,tohyama99,tohyama01}.
We first consider possible effects of superconductivity and antiferromagnetism 
on our phase diagram, which are not taken into account in our leading order theory. 
Typically, superconductivity in $e$-cuprates occurs 
below 25 K \cite{armitage10}, which is around 
$0.004 t$ in the current theory for $t\sim 500$ meV \cite{hybertsen90}.
Because our charge-order instabilities occur higher than this temperature, 
a major part of our results could not be affected by superconductivity, although 
charge orders or their tendencies would be suppressed inside the superconducting state. 
On the other hand, antiferromagnetism extends to the region of $11$-$14\%$ doping 
in $e$-cuprates \cite{armitage10}.  Because charge orders occur 
below $13$-$14\%$ in our phase diagram, most of charge orders 
that we have found could be masked by antiferromagnetism; yet they can be 
observed in Pr$_{1-x}$LaCe$_{x}$CuO$_{4}$, which has a lower 
critical doping rate of antiferromagnetism. 
The most relevant charge orders in a moderate doping region  are 
various BOPs with $\vq$ close to $(\pi,\pi)$, which 
become dominant below $T\sim 0.01t \sim 50$ K, as seen in \fig{fig4}. 
Figure~\ref{fig4} also implies that the $d$CDW can become relevant to $e$-cuprates 
if the critical temperature of antiferromagnetism becomes 
lower than $T_c$ of the $d$CDW in a certain doping region. 
At present, experimental evidence of neither BOP nor $d$CDW is obtained 
in $e$-cuprates. However, given that evidence of some order 
competing with superconductivity was obtained quite recently 
in $e$-cuprates \cite{hinton13} and 
that a new type of charge order was also found quite recently in 
$h$-cuprates \cite{wu11,ghiringhelli12,chang12,comin14,neto14},
it may be too early to reach a conclusion about a possible 
charge instability in $e$-cuprates. 
In particular, BOPs with $\vq$ close to $(\pi,\pi)$ are not reported 
in $h$-cuprates, and thus in this sense $e$-cuprates 
are attractive for exploring a new type of charge order in cuprates. 
Even if charge-order instability does not occur, 
its fluctuation effect can be observed as collective excitations. 
It is interesting to explore a possible connection with the new collective mode recently found 
in optimal $e$-cuprates by resonant inelastic X-ray scattering \cite{wslee13}.

There is growing evidence that the PG 
is related to some charge order or its fluctuations in $h$-cuprates \cite{tanaka06,vishik10,kondo11,yoshida12,hinkov08,daou10}.
In particular, the $d$CDW \cite{chakravarty01,greco09,bejas11,greco11} 
and $d$PI \cite{yamase09,hackl09,yamase12} are candidates. 
If the charge order is indeed responsible for the PG, 
the current theory suggests that the property of the PG 
should be different between hole doping and electron doping because of 
the strong particle-hole asymmetry of charge-order instabilities. 
Although a PG was reported in the optical conductivity spectra in the non-superconducting 
crystals of $e$-cuprates \cite{onose01},
the PG corresponding to the PG observed in 
$h$-cuprates, namely in a doping region where the superconducting phase 
occurs at low $T$, seems to be missing or at least much weaker. 
It is quite interesting to study whether 
other scenarios of the PG such as fluctuations associated with 
Cooper pairing and antiferromagnetism can provide a natural explanation 
of the asymmetry of the PG between hole-doped and electron-doped  
cuprate superconductors.

\section {Conclusions} \label{sec:conclusions}
We have performed a stability analysis of the paramagnetic phase in the 
two-dimensional $t$-$t'$-$J$ model 
by employing a leading order theory formulated in a large-$N$ expansion scheme. 
Our theoretical framework has the advantage of taking into account
all possible charge instabilities on equal footing 
and of allowing us to perform a comprehensive study of charge instabilities 
in a controllable scheme. 
We have taken a positive value of $t'$ and our results can be relevant to $e$-cuprates. 
To the best of our knowledge, no systematic studies of charge instabilities 
have been performed for $e$-cuprates, even in the large-$N$ expansion. 
We have found that the $d$CDW and $d$PI become relevant rather close to 
half filling and that various types of BOPs with $\vq$ close to $(\pi,\pi)$ 
are dominant in a moderate doping region. 
PS is also enhanced but can be suppressed substantially by 
the nearest-neighbor Coulomb repulsion $V$, although the instabilities associated with 
BOPs, the $d$CDW, and the $d$PI are almost intact even in the presence of large $V$. 

The charge order tendencies we have found for $t'>0$ are 
very different from those for $t'<0$ \cite{bejas12}.
This strong particle-hole asymmetry implies that charge orders 
are less favorable in $e$-cuprates, although they can still occur. 
Furthermore, if charge orders are responsible for the PG, 
the current theory may naturally explain the reason why the PG phenomenon 
is very different between $e$-cuprates and $h$-cuprates.

\ack
The authors thank A M Oles for a critical reading of the manuscript.
AG thanks the National Institute for Materials Science (NIMS), 
where this work was initiated, and the Max Planck Institute for hospitality.
HY was supported by a Grant-in-Aid for Scientific Research from Monkasho 
and the Alexander von Humboldt Foundation.

\section*{References}
\bibliography{main_pc.bib}
\end{document}